\documentclass[twocolumn,twoside,preprintnumbers,amsmath,amssymb,showkeys]{revtex4}
\usepackage{epsfig}
\usepackage{graphicx}% Include figure files

\usepackage{fancyhdr}

\usepackage{pslatex}

\begin{document}

\title{Numerical symmetrization of state of identical particles}

\author{Oleg Utyuzh$^1$, Grzegorz Wilk$^1$ and Zbigniew W\l odarczyk$^2$}

\affiliation{$^1$ The Andrzej So\l tan Institute for Nuclear
                  Studies, Ho\.za 69, 00-681 Warsaw, Poland \\
             $^2$ Institute of Physics, \'Swi\c{e}tokrzyska Academy,
                  \'Swi\c{e}tokrzyska 15, 25-405 Kielce, Poland}

%\received{on ????????????, 2006}

\begin{abstract}

The method of numerical symmetrization of state of identical
particles proposed by us before is clarified and discussed.


\end{abstract}
\maketitle

\thispagestyle{fancy}

\setcounter{page}{1}

\section{Introduction}

It is well known experimental fact that identical pions, which are
produced in heavy ion collisions, being bosons show Bose-Einstein
correlations. These correlations result from the quantum mechanical
interference in the corresponding symmetrized $n$-particle wave function
(where $n$ denotes the number of produced identical pions). They contain
large amount of information about the statistical properties of the
momentum and configuration space distribution of the system, and thus
provide a potentially very useful method to probe the geometry of the
hadronizing source (see, for example, \cite{General} and references
therein).

According to common understanding we are not able to determine which pion
is emitted from which position in the source, so we are required by Bose
statistics to add amplitudes for all possible alternate histories. In
general, symmetrized wave function for $n_{\pi}$-pion state can be
written in the following way \cite{zajc}:
\begin{equation}
\Psi_{\left\{p\right\}}(\left\{x\right\})=\frac{1}{\sqrt{n_{\pi}!}} \,
\sum_{\sigma}\exp\left[-i\sum_{j=1}^{n_{\pi}} p_j\, r_{\sigma(j)}\right]
,\label{eq:nwave}
\end{equation}
where $\sigma(j)$ denotes the $j^{th}$ element of a permutation of the
sequence $\{1,2,...,n_{\pi}\}$ and the sum over $\sigma$ denotes
therefore the sum over all $n_{\pi}!$ permutations in this sequence
(dependence on positions of points of detection, which will vanish when
calculating probabilities, was neglected). Here $r$'s denote the points
of production of secondaries. Because in the experiment one observes only
momenta of produced secondaries these $r$'s  must be somehow get rid of.
This is so far always done by integrating over $\{r_j\}$ with some {\it
assumed} distribution $\rho(\{r_j\})$, which is {\it assumed} to be
factorizable and expressed by product of independent single particle
distributions $\rho(\{r_j\})=\prod_j \rho(r_j)$ \cite{General,zajc}. As
result one gets the following probability of the $n_{\pi}$-pion state,
\begin{eqnarray}
\!\!\!\! \mathcal{P}_{1,..,n_{\pi}}\!\!\!\!\!&=&\!\!\!\!\!
\frac{1}{n_{\pi}!}\sum_{\sigma}\prod_{j} \Phi_{j,\sigma(j)}\equiv
\frac{1}{n_{\pi}!}
\begin{tabular}{||cccccc||}
& $\Phi_{1,1}$ & & $\cdots$ & & $\Phi_{1,n_{\pi}}$ \\
& $\vdots$ & & $\Phi_{j,j}$ & & $\vdots$ \\
& $\Phi_{n_{\pi},1}$ & & $\cdots$ & & $\Phi_{n_{\pi},n_{\pi}}$ \\
\end{tabular} \label{eq:permanent}
\end{eqnarray}
expressed by {\it permanent} of the matrix $||\Phi_{ij}||$, where
\begin{equation}
\Phi_{ij}=\int e^{i(p_i-p_j)r}  \rho(\{r\}) d^4 \{r\} .
\end{equation}
For usually considered $n_{\pi}=2$ case one recovers well known
classical expression for the probability of detecting two pions in
the final state \cite{General}:
\begin{equation}
\mathcal{P}_{1,2} = \frac{1}{2}
\begin{tabular}{||ccc||}
& $\Phi_{1,1}$ & $\Phi_{1,2}$ \\
& $\Phi_{2,1}$ & $\Phi_{2,2}$
\end{tabular}
=\frac{1}{2}\left( \Phi_{1,1}\Phi_{2,2}+\Phi_{1,2}\Phi_{2,1}\right) ~.
\end{equation}
Unfortunately the execution time of direct computation of the permanent,
eq.(\ref{eq:nwave}), grows exponentially with $n_{\pi}$ one has to devise
some special methods like {\it Metropolis procedure} investigated in
\cite{zajc,Germans} or {\it von Neumann accepting-rejection method}
proposed in \cite{Cramer}.

The {\it Metropolis procedure} proposed in \cite{zajc} uses the standard
Monte Carlo technique due to Metropolis. This is general method which
allows to generate ensemble of $n$-body configurations according to some
prescribed probability density. That is, the probability of a given
configuration in the ensemble is precisely that given by the probability
density used to generate "successive" configurations. In \cite{zajc} this
technique was used to modify directions of momentum vectors of number of
selected particles from a system of $n$ identical particles in order to
impose the $n$-particle distributions derived from BE correlation
functions. This procedure is then repeated many times, changing selected
particles, until a kind of "equilibrium" is achieved. As shown in
\cite{zajc} one was able in this way to generate typical multipion events
which explicitly exhibit all correlations induced by Bose statistics. As
a result of application of this algorithm a number of objects (called
{\sl speckles}) being clusters of large number of identical pions in the
phase-space is being formed. The drawback of this method is that
symmetrization of clusters with sizes (represented by the number of
particles inside cluster, $n_{cluster}$) larger than $n_{cluster}\approx
10$ takes prohibitively long time. In \cite{Germans} this method was
therefore modified by limiting symmetrization only to particles in
clusters. This was possible by using wave packets to describe produced
particles instead of plane waves used in \cite{zajc} allowing therefore
for localization of particles within certain phase-space volume and for
providing the suitable criterium for defining a cluster.

The {\it accepting-rejection method} investigated in \cite{Cramer} is
based on the well known "hit-or-miss" technique of generating a set of
random numbers according to a prescribed distribution. The method was
designed to model collapse of a multiparticle wave function into a
properly symmetrized state, as required by Bose quantum statistics. In
contrast to the previous one it is sequential because $n$ multiparticle
event is constructed by first choosing single particle in phase space,
then adding to it the second one according to the assumed $2$-particle
correlation function $C_2\propto P_2$, then adding $3^{th}$ particle
according to $C_3 \propto P_3$ and so on. It is easy to realize that in
this way one gets in the allowed phase space a "cell-like" structure
because regions with some particles inside them already present will have
bigger chance to attract new particle. In a sense it looks like follows:
first particle forms a seed for a first cell. When second particle is
added to event it can, depending on its distance from the first one,
either remain in that cell or later on attract new particles and in this
way start to form a new cell. This will then continue until all particles
are used. Unfortunately, this sequential procedure is even more time
consuming than the previous one.

\section{Our proposition}

The above discussion shows that complexity of numerical symmetrization of
wave function for all identical pions produced in a given event (cf.,
eq.(\ref{eq:nwave})) can be substantially reduced if only one can justify
the idea that such symmetrization should be applied to groups of limited
number of particles, as proposed in \cite{Germans}. As argued there this
can be achieved by dividing, according to some prescribed procedure, the
initial set of secondaries into clusters consisting of identical
particles with similar momenta. Such groups of particles, but this time
of (almost) equal energies, were introduced in \cite{EEC} under the name
{\it Elementary Emitting Cells} (EEC's). The physical justification was
that bosonic particles, because of their statistical properties, tend to
occur, as much as possible, in {\it the same} state. One should realize
now that such decomposition corresponds to {\it factorization} of
permanent given by eq.(\ref{eq:permanent}) into matrix with a block
structure:
\begin{equation}
\begin{tabular}{||cccccc||}
& $\Phi_{1,1}$ & & $\cdots$ & & $\Phi_{1,n_{\pi}}$ \\
& $\vdots$ & & $\Phi_{i,i}$ & & $\vdots$ \\
& $\Phi_{n_{\pi},1}$ & & $\cdots$ & & $\Phi_{n_{\pi},n_{\pi}}$ \\
\end{tabular}
\,\,\Rrightarrow
\begin{tabular}{||cccccc||}
& \fbox{$EEC_1$} & & $\cdots$ & & 0 \\
& $\vdots$ & & $\ddots$ & & $\vdots$ \\
& 0 & & $\cdots$ & & \fbox{$EEC_{N_{cell}}$}\, \\
\end{tabular} \,\,\, .
\end{equation}

This idea of EEC was exploited by us further in \cite{EEC1} where instead
of symmetrization of multiparticle wave function (depending on space-time
positions, $x$, and energy-momenta, $p$) we worked in the number of
particles basis. The bosonic character of the produced secondaries means
in this case the specific bunching of identical particles in the phase
space. In fact it is nothing else but modelling the correlations of
fluctuations due to quantum statistics present in the system, which for
$2$-particle case are represented by:
\begin{equation}
\langle n_1 n_2\rangle =
                \langle n_1\rangle \langle n_2\rangle
                 + \rho \sigma(n_1)\sigma(n_2) \label{eq:COV}
\end{equation}
$\sigma(n)$ is dispersion of the multiplicity distribution $P(n)$
and $\rho$ is the correlation coefficient depending on the type of
particles produced: $\rho = +1,-1,0$ for bosons, fermions and
Boltzmann statistics, respectively). The important feature of the
EEC method is that number of particles in each cell follows by
definition the geometrical (or Bose-Einstein) distribution
characterized for identical bosonic particles.

It should be mentioned at this point that the importance of bunching of
particles in modelling quantum statistical effects has been demonstrated
already in \cite{MIE}. In this paper, following ideas of information
theory, authors constructed Monte-Carlo (MC) event generator for
multiparticle production processes (and applied it to $e^+e^-$
reactions). The main point was the assumption that particles of the same
charge are located in cells (in their case they were constructed in
rapidity space and were of equal size). It turned out that such division
of phase space into cells was crucial for obtaining the characteristic
form of the $2$-body BEC correlation function $C_2(Q=|p_i - p_j|)$ (i.e.,
the one peaked and greater than unity at $Q = 0$ and then decreasing in a
characteristic way towards $C_2 = 1$ for large values of $Q$, out of
which one usually deduces the spatio-temporal characteristics of the
hadronization source \cite{General}).

\begin{figure}[!htb]
%\vspace*{-1.cm}
\begin{center}
\includegraphics*[width=8cm]{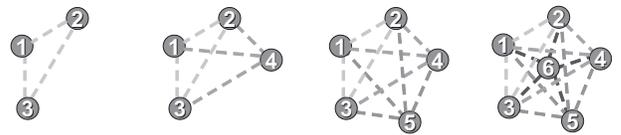}
\end{center}
\vspace*{-0.6cm} \caption{Representation of algorithm for the sequential
model, eq. (\ref{sequential}).} \label{fig2} \vspace{-0.3cm}
\end{figure}

In our case we argue that the method of production of EEC proposed in
\cite{EEC} can be used to {\it define} the structure of clusters obtained
in \cite{Germans} (with, as it turns out, about $n_{cl}=2$ particles per
cluster on average, depending on circumstances \cite{ours}). Therefore,
instead of symmetrizing all particles in a given event, one symmetrizes
separately particles in a number of EEC's with $n_{cl}\ll n_{\pi}$. In
the plane wave approximation used in \cite{zajc}, one has therefore for
some typical EEC the following $n_{cl}$-particle probability function:
\begin{equation}
\mathcal{P}_{1,...,n_{cl}}^{\left[n\right]}=1+\frac{2}{n_{cl}!}\sum_{\sigma'>\sigma=1}^{n_{cl}!}
\cos\left\{\sum_{j=1}^{n_{cl}}p_i\left[r_{\sigma(j)}-r_{\sigma'(j)}\right]\right\}
. \label{full}
\end{equation}
Notice that eq. (\ref{full}) still contains (unmeasurable) positions of
production of particles, $\{r_{\sigma(j)}\}$. Therefore they will be
later eliminated by selecting them from some {\it assumed} distribution
in a kind of numerical integration process. It corresponds to analytical
integrations encountered before but in our approach we do not limit in
any way the form of density function used (it had to be {\it
factorizable} before). It should be stressed that when all particles in
the cluster are exactly in the same state then one gets, as expected,
\begin{equation}
\mathcal{P}_{1,...,n_{cl}}^{\left[n\right]}\left({\rm
max}\right)\left|_{p_1=\cdots=p_{n_{cl}}} \right.= n_{cl}!
.\label{max}
\end{equation}
However, in practice, even now from time to time one encounters EEC
with $n_{cl}\gg 1$, in which case eq.(\ref{full}) is still very time
consuming. For such cases we have to use some approximate schemes.
The first one is sequential (cf. FIG.~\ref{fig2}): one starts with
some single particle in EEC and adds new particles one by one to all
others already present and correlate them by using standard
$2$-particle probability: $1+\cos\left(\delta r\cdot\delta
p\right)$. The resulting probability to get $n_{cl}$-particle state
is given by:
\begin{equation}
\mathcal{P}_{1,...,n_{cl}}^{\left[2\right]}=\mathcal{P}_{1,...,n_{cl}-1}\cdot
\prod_{j=1}^{n_{cl}-1}\left[1+\cos\left(\delta
r_{jn_{cl}}\cdot\delta p_{jn_{cl}}\right)\right] .\label{sequential}
\end{equation}

\begin{figure}[!htb]
%\vspace*{-1.cm}
\begin{center}
\includegraphics*[width=7.5cm]{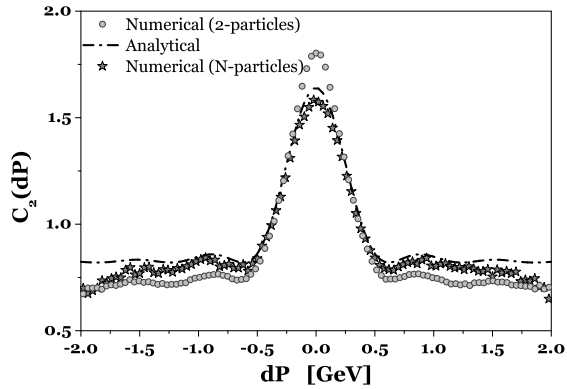}
\end{center}
\vspace*{-0.6cm} \caption{Example of $C_2$ modelled for single EEC with
$N=4$ by using $2$-particles (sequential) correlations as given by eq.
(\ref{sequential}) (circles) and full $N$-particles correlations as given
by eq. (\ref{full}) (stars). For comparison, the analytical result,
$C_2(\delta P) = 1 + \sin^2 (R\delta P)/(R\delta P)^2$, obtained for
uniform one dimensional source in space with $R=1$ fm is also shown
(dotted-dashed line).} \label{fig1} \vspace{-0.5cm}
\end{figure}
\begin{figure}[!hbt]
\vspace*{0.2cm}
\begin{center}
\includegraphics*[width=7.5cm]{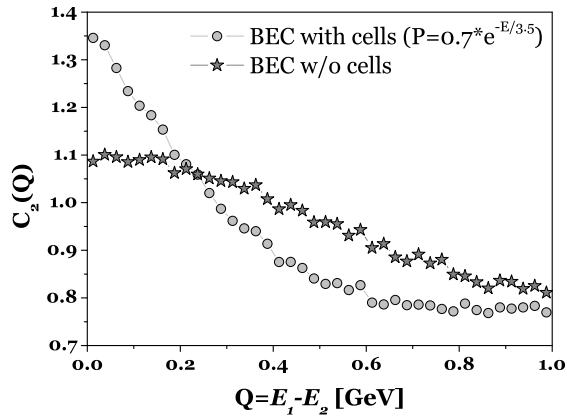}
\end{center}
\vspace*{-0.6cm} \caption{Comparison of $C_2$ modelled by using MC event
generators with EECs (circles) and without EECs but with particles
selected directly from the corresponding Bose-Einstein distribution
\cite{THERMINATOR}. In both cases as reference event the Boltzmann
distribution was used.} \label{fig3} \vspace{-0.5cm}
\end{figure}
Although appealing (in fact it resembles procedure used in \cite{Cramer})
it has drawback that when all particles are in the same state then
\begin{equation}
\mathcal{P}_{1,...,n_{cl}}^{\left[2\right]}\left({\rm max
}\right)\left|_{p_1=\cdots=p_{n_{cl}}} \right. =
2^{\frac{1}{2}n_{cl}(n_{cl}-1)}
\end{equation}
i.e., for $n_{cl} >2$ the correlations are stronger and the maximal
value of probability function gets bigger than allowed limit defined
by eq.(\ref{max}). Both, exact (\ref{full}) and sequential
(\ref{sequential}) methods are compared in FIG. \ref{fig1}.
Approximation (\ref{sequential}) keeps the width of $C_2$ the same,
however it differs substantially in $C_2(\delta P=0)$ (from which
one tends to estimate the so called {\it chaoticity} of the
hadronizing source \cite{General}). The other possible
approximation, which was used in our hitherto applications
\cite{ours}, is that all particles in a given EEC are correlated
only with the first particle defining this cell, not between
themselves. It is interesting to notice that the results for $C_2$
obtained this way are almost the same as those obtained by using the
full method, eq. (\ref{full}), but differ from that of eq.
(\ref{sequential}).

We close this section by noticing recent attempt to imitate the bosonic
nature of particles produced by MC event generator, in which they were
choosen directly according to Bose-Einstein distribution
\cite{THERMINATOR}. This method seems to be very natural and occurred to
be also very fast. However, as one can see from FIG.~\ref{fig3}, where it
is compared with our approach, it leads to completely different $C_2$
substantially underestimating the correlation function $C_2$. We claim
therefore that such procedure inevitably loses some piece of important
information, namely the fact that EEC's are formed and that Bose-Einstein
distributions are only for particles in such EEC's, not in the whole
event. In fact, signal of possible Bose-Einstein correlations without
cells is seen in FIG. \ref{fig3} are the trivial correlations, which can
be eliminated by the proper choice of the reference event.
\vspace{-0.01cm}

\section{Summary}

We would like to summarize by stressing that there is {\it no way} to add
to any of the existing MC event generators effects of quantum statistics,
in particular Bose-Einstein one (BE). This is because they are all build
on basis of classical physics with both the space-time and
energy-momentum characteristics of produced secondaries used
simultaneously. The only way out advocated here (albeit, most probably,
not very practical one and therefore hardly to be followed) is to build
the multiparticle MC event generator {\it ab initio}, with BE properties
(like bunching in phase space) being one of its basic principles and
consisting its first step. All other features of such generator would
have to be added only after this. So far there is only one working
example of this type \cite{MIE}, our efforts \cite{ours} aim for its
further generalization and will be continued.

\noindent{\bf Acknowledgements}

OU is grateful for support and for the warm hospitality extended to
him by organizers of the ISMD2006.  Partial support of the Ministry
of Science and Higher Education (grants Nr
621/E-78/SPB/CERN/P-03/DWM 52/2004-2006 and 1 P03B 022 30) is
acknowledged (OU and GW).
%The subject reviewed here has been investigated in
%collaboration with G.Wilk and Z.W\l odarczyk.

%\smallskip


\begin{thebibliography}{99}

\bibitem{General} T.J Humanic, Int.J.Mod.Phys. {\bf E15}, 197-236 (2006).

\bibitem{zajc} W.A. Zajc, Phys. Rev. {\bf D35}, 3396 (1987).

\bibitem{Germans} H. Merlitz and D. Pelte, Z.Phys. A {\bf 351},187 (1995);
Z. Phys. A {\bf 357}, 175 (1997) .

\bibitem{Cramer} J.~G.~Cramer, {\it Event
Simulation of High-Order Bose-Einstein and Coulomb Correlations},
Univ. of Washington preprint (1996, unpublished).

\bibitem{EEC} M.Biyajima, N.Suzuki, G.Wilk and Z.W\l odarczyk;
              Phys. Lett. B {\bf 386} (1996) 297.

\bibitem{EEC1} O.V.Utyuzh, G.Wilk and Z.W\l odarczyk, Phys. Lett. B
{\bf 522}, 273 (2001) and Acta Phys. Polon. B {\bf 33}, 2681 (2002)
(hep-ph/0205087).

\bibitem{MIE} T. Osada, M. Maruyama and F. Takagi, Phys. Rev. D {\bf 59} (1999) 014024.

\bibitem{ours} O.V.Utyuzh, G.Wilk and Z.W\l odarczyk,
               hep-ph/0503046,  Acta Phys. Hung. {\bf A} -
               Heavy Ion Phys.{\bf 25} (1) (2006) 83;
               hep-ph/0509320, AIP Conf. Proc. {\bf 828} (2006) 75;
               hep-ph/0509342, to be published in Nukleonika {\bf
               51} (Supplement 3) (2006).

\bibitem{THERMINATOR} A. Kisiel, T. Ta\l u\'{c}, W. Broniowski and
W. Florkowski, Comp. Phys. Comm. {\bf 174}, 669 (2006).

\end{thebibliography}
\end{document}